\documentclass[]{spie}  
\usepackage[]{graphicx}
\usepackage[francais]{babel}
  \usepackage[applemac]{inputenc} 
  \usepackage[]{amsmath} 
  \usepackage{rotfloat}
  \usepackage{float}
  \usepackage{hyperref}
  \usepackage{placeins}

\title{Some results after 10 years of site testing at Concordia, Antarctica}

\author{Eric Fossat\supit{a}.
\skiplinehalf
\supit{a}  Université de Nice Sophia-Antipolis, Centre National de la Recherche Scientifique , Observatoire de la Côte d'Azur, UMR 6525 H. Fizeau,Campus Valrose, F-06108 Nice cedex, France.\\
}
\authorinfo{Further author information: (Send correspondence to E. Fossat)\\E-mail: eric.fossat@unice.fr}
 
\begin{document}  
 \maketitle
\begin{abstract}

At an altitude of 3250m and at a latitude of 75°S, the Italo-French Concordia station was open to winter-over teams in 2005. It is one of the high points of the Antarctic polar plateau. These extreme remote sites are expected to provide exceptional conditions for astronomical observations, specially in the infra-red ranges, given the very cold winter temperatures, averaging well below -60C. Being very flat as highest points of that very broad polar plateau, they are also not subject to the famous katabatic winds that can be devastating on the Antarctic coast, and in fact their mean wind speed along the year are the weakest known on Earth, less than 3 m/s. Besides the resulting absence of danger that such winds would present for large size optical instruments, this situation offers another benefit, which is an excellent free atmosphere seeing above a very thin but turbulent surface layer. This paper emphasizes these seeing peculiarities, but not only. It is presented as simply following a significant fraction of my slide presentation during the meeting.

\end{abstract}

%\keywords{High angular resolution, Optical Interferometry, Direct Imaging, Coronagraphy, Achromatic, Exoplanet.}

\section{Introduction}
The highest points of the Antarctica plateau have many obvious astronomical advantages over competing sites, due to their climate and remoteness from any polluting civilization. Among these obvious points are the very long continuous observations made possible by the polar day or the polar night. Another one is the extremely cold and dry atmosphere which is favoring infra-red observations. The night time Precipitable Water PW is averaging at or below 0.2 mm (Tomasi et al, 2006), thus opening windows in the near infra-red range, i.e. around K-dark and between 3 and 5 microns, and other interesting windows in the TeraHz range, especially at 350 and 200 microns where the benefit is of the order of a factor 5 against the high altitude Chilean sites. Less obviously, they also benefit from an interestingly unique distribution of turbulence. This has been extensively measured at Dome C since the first winter-over authorized in 2005 by the Franco - Italian Concordia station operation (Aristidi et al, 2009). During winter and summer very different but both unusual vertical distributions of the turbulent energy are measured. The situation is in general dominated by a surface inversion layer that becomes very turbulent when the temperature gradient is steep in winter, and can completely vanish in summer when this gradient becomes flat. In summer the situation depends on the Sun's elevation, and is then strongly time dependent, having an optimum period of a few hours of excellent seeing every day in the middle of local afternoon (Aristidi et al, 2005). In the other 3 seasons, the mean seeing is almost only altitude dependent above the snow surface. The turbulent layer contains, statistically, 95 percent of the total C$n^2$ along the line of sight. Its geometrical properties are statistically independent of the season, within the measurement accuracy. Above this layer, the mean seeing value is also found independent of the season, between 0.3 and 0.4 arc-sec as soon as the telescope is located above a sharply defined altitude threshold, which fluctuates about a mean value of the order of 25 m. The non summer seeing therefore exhibits a nearly bimodal statistical distribution. It is indeed as good as around 0.3 arc-sec 50 percent of the time 25m above the surface, this fraction of time decreasing to about 40  percent at 20m and slightly less than 20 percent at 8m. But it is obviously not equivalent to have 40 percent of the good seeing distributed in many short periods of from seconds to minutes rather than in extended long sequences of hours or days. This paper also addresses the temporal distribution of this good seeing percentage. That part extends the first analysis of Aristidi et al. (2009), by applying a method for correcting the effects of the gaps in the data.

\begin{figure}[htbp]
\centering
 \includegraphics[width=10cm]{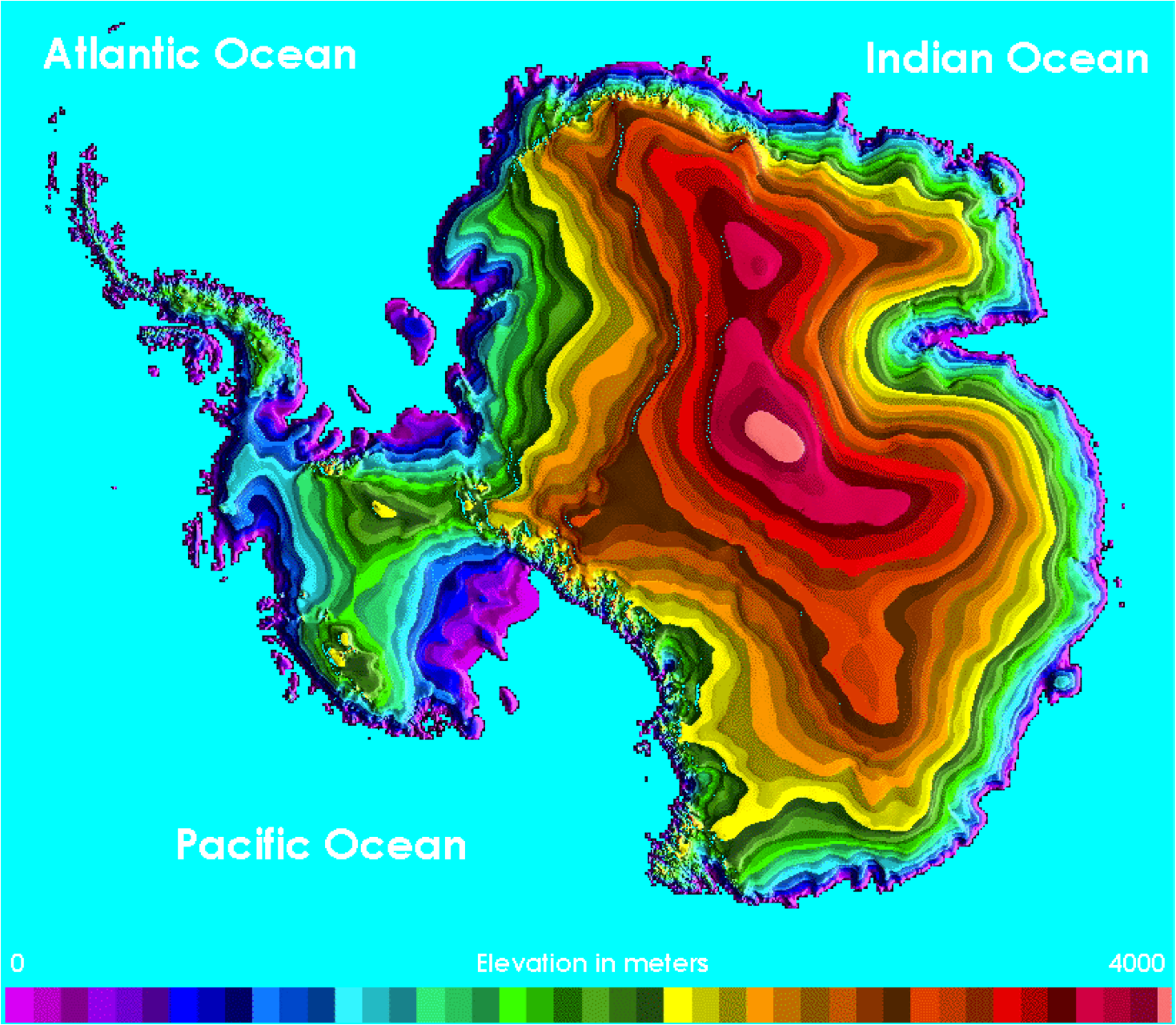}
 \caption[]{The Antarctic continent is larger than Europe and includes the Geographical South Pole, which is the central point of this square figure. It is totally covered by ice, its maximum thickness being more than 4000m  at Dome A, the central summit on the figure, and still above 3600 m at Dome F (the upper one) and more than 3200m at Dome C (the lowest one).}
 \label{monobloc}
   \end{figure} 
   
   \begin{figure}[htbp]
\centering
 \includegraphics[width=10cm]{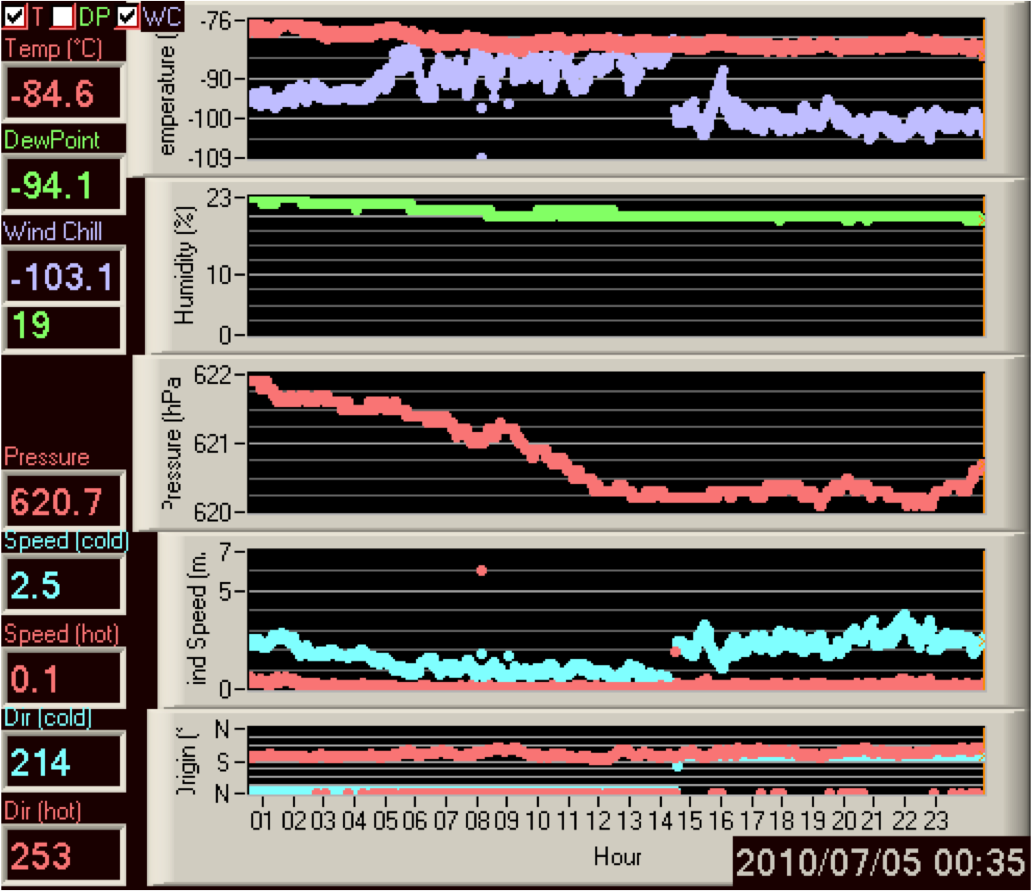}
 \caption[]{The winter temperatures are very cold at Dome C. This is a one-day meteorological record from July, 2010 that shows, from top to bottom: temperature and wind chill (Celsius), relative humidity, atmospheric pressure (hPa), wind speed (m/s) and wind direction. That is a typical good situation, when the wind is oriented from the South, thus coming from the central, coldest and dryiest part of the continent. This is the prevailing wind direction at Dome C. Note that the wind speed is mostly comprised between 0 and 3 m/s and that the atmospheric pressure, around 620 hPa, correspond to values encountered at 4000 m of altitude at mid latitudes, despite the local altitude of only 3250m. This is due to the thermodynamical property of the very cold air, and is another good point for astronomers }
 \label{monobloc}
   \end{figure} 
   
    \begin{figure}[htbp]
\centering
 \includegraphics[width=9cm]{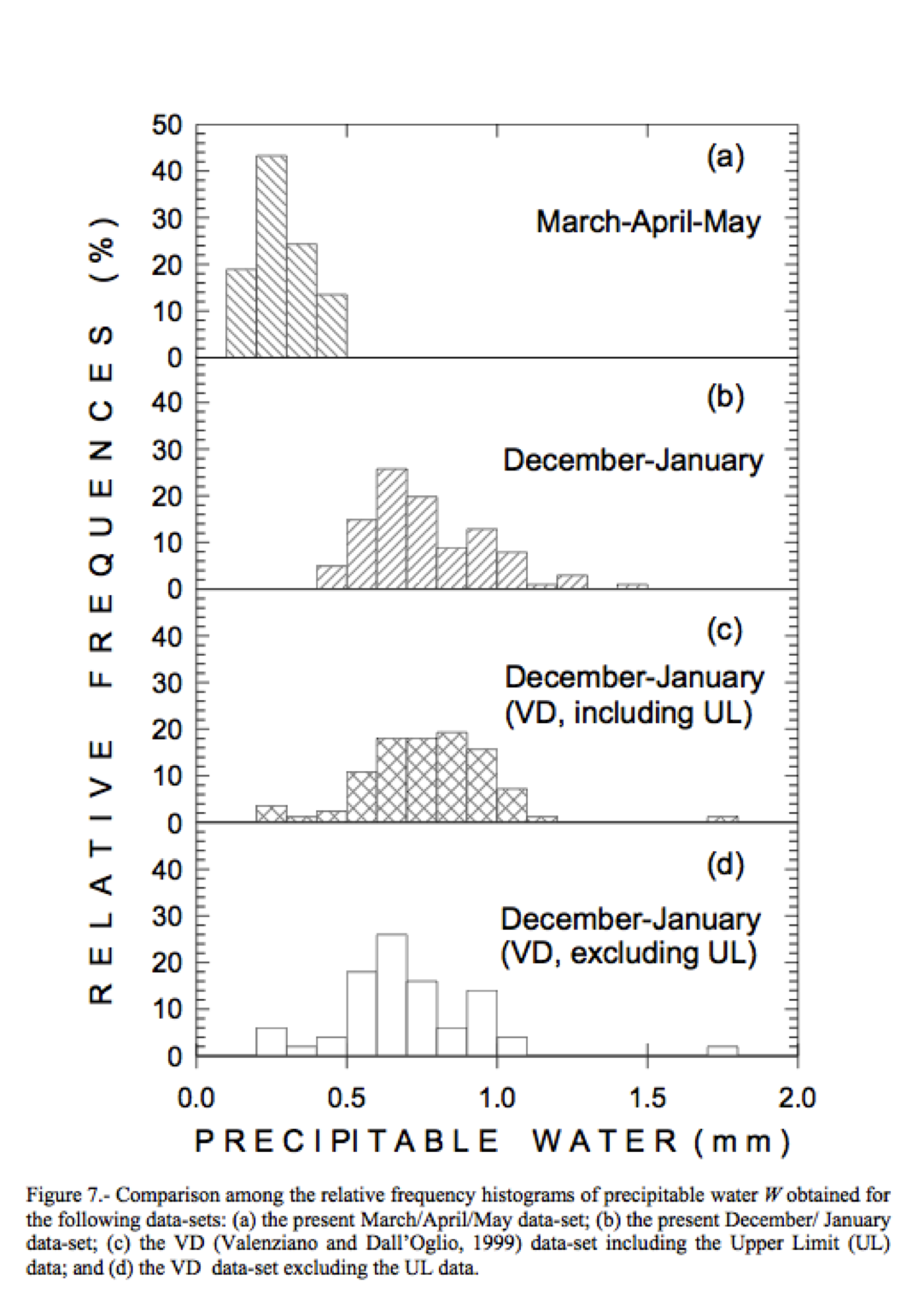}
 \caption[]{Figure from Tomasi et al, 2006. The PW has been measured in local summer (b, c, and d, December - January) and Autumn and early winter (a, March - May). The deep winter values (May, June, July) are missing, but they can only be a little dryer }
 \label{monobloc}
   \end{figure}

    \begin{figure}[htbp]
\centering
 \includegraphics[width=10cm]{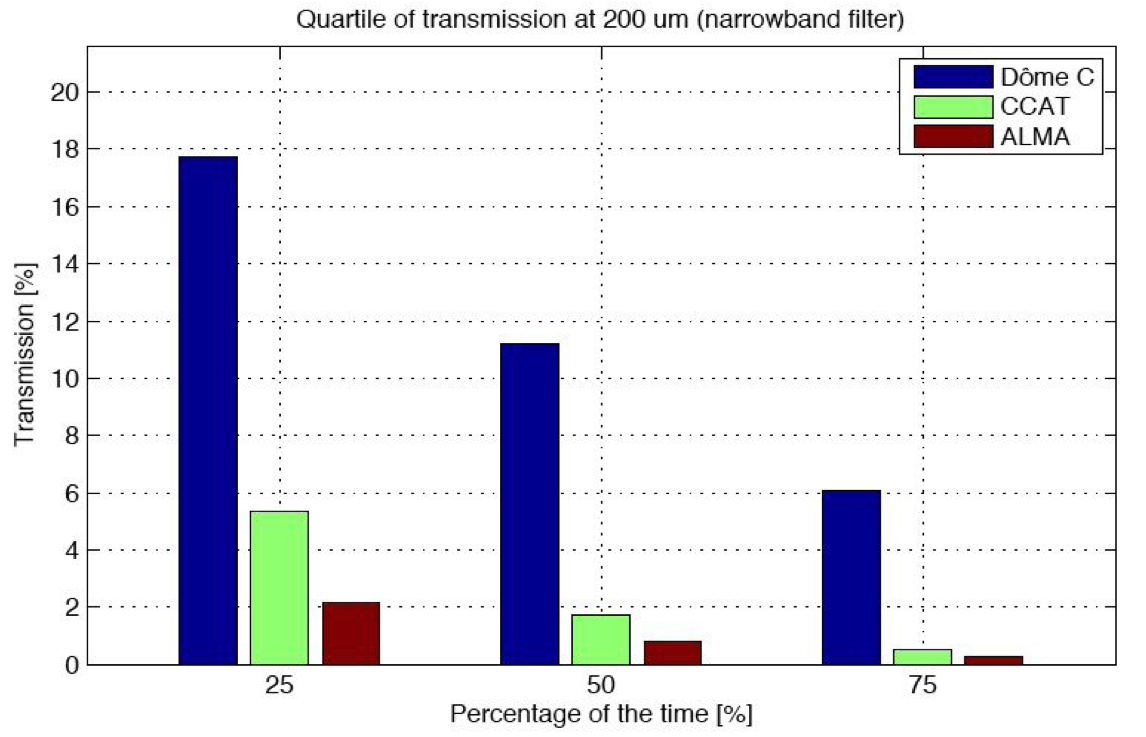}
 \caption[]{Figure from xxxxxx. The extremely dry Dome C atmosphere in winter opens a spectral window at 200 microns that is still not to be compared to space, but offers a transmission 5 times better than the high  altitude Chilean sites, making its exploitation realistic. And of course, there are other infra-red windows, for instance between 2 and 5 microns, that also benefit very significantly from this exceptionally dry conditions.  }
 \label{monobloc}
   \end{figure}       
   
    \begin{figure}[htbp]
\centering
 \includegraphics[width=11cm]{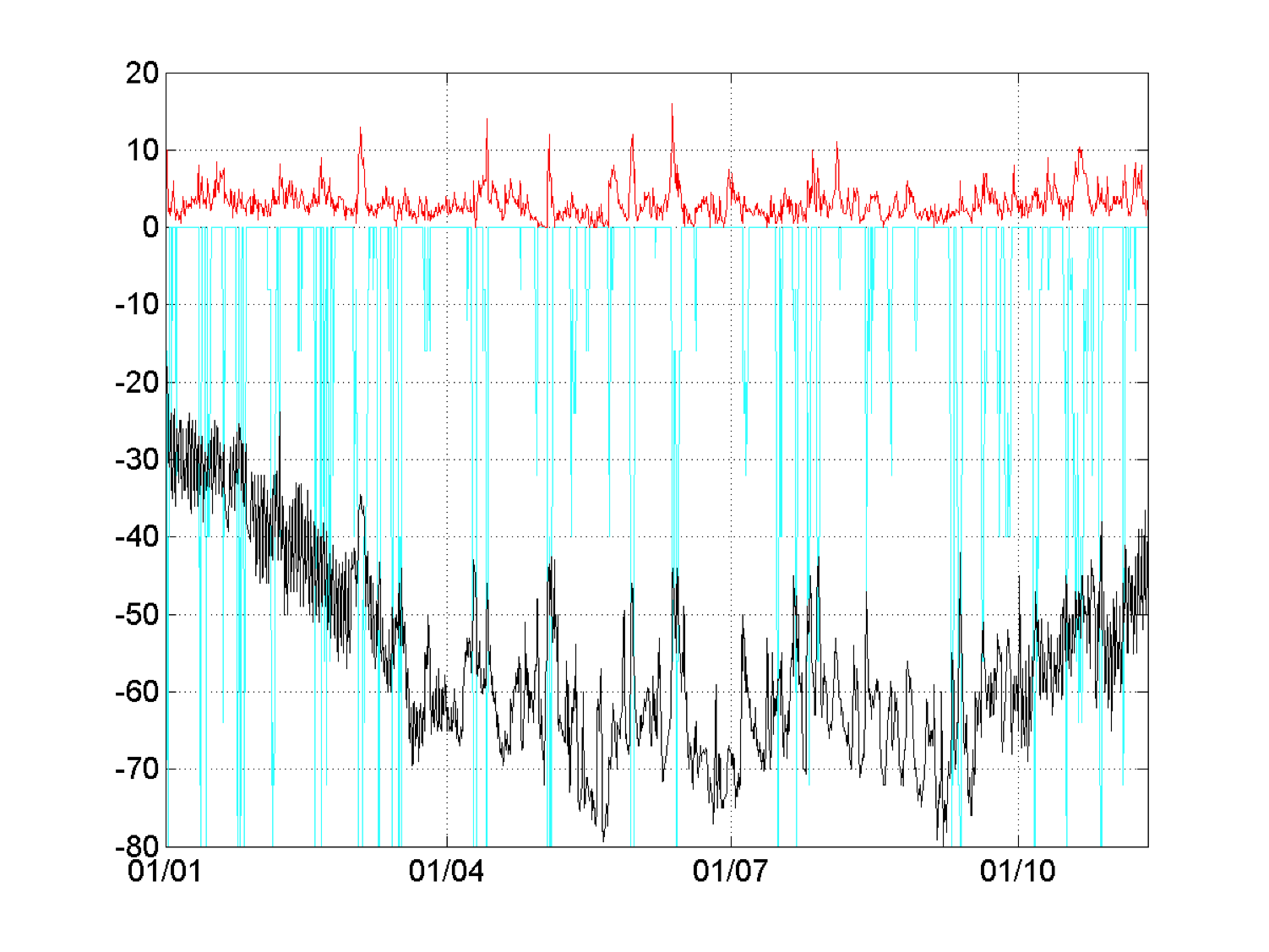}
 \caption[]{Figure prepared by Schmider, private comm. Clear sky fraction (light blue), surface wind speed (red) and temperature (black) recorded during one year, in 2006 (second winter-over). Wind speed and temperatures are those from the Concordia AWS, by courtesy of L. Agnoletto and A. Pellegrini. The clear sky fraction has been estimated by naked eye by E. Aristidi (Mosser and Aristidi, 2007). It is very clear that the bad weather episodes, specially in winter, are visible simultaneously on the three parameters, so that the 10 to 15 percent of time lost because of too strong winds, of too warm temperature or of cloudy sky are not adding up to 30 or 40 percent, but to still only 10 to 15 percent. A good new for the long integrations needed by all the photometric or spectroscopic time series analyses to come.  }
 \label{monobloc}
   \end{figure}

    \begin{figure}[htbp]
\centering
 \includegraphics[width=11cm]{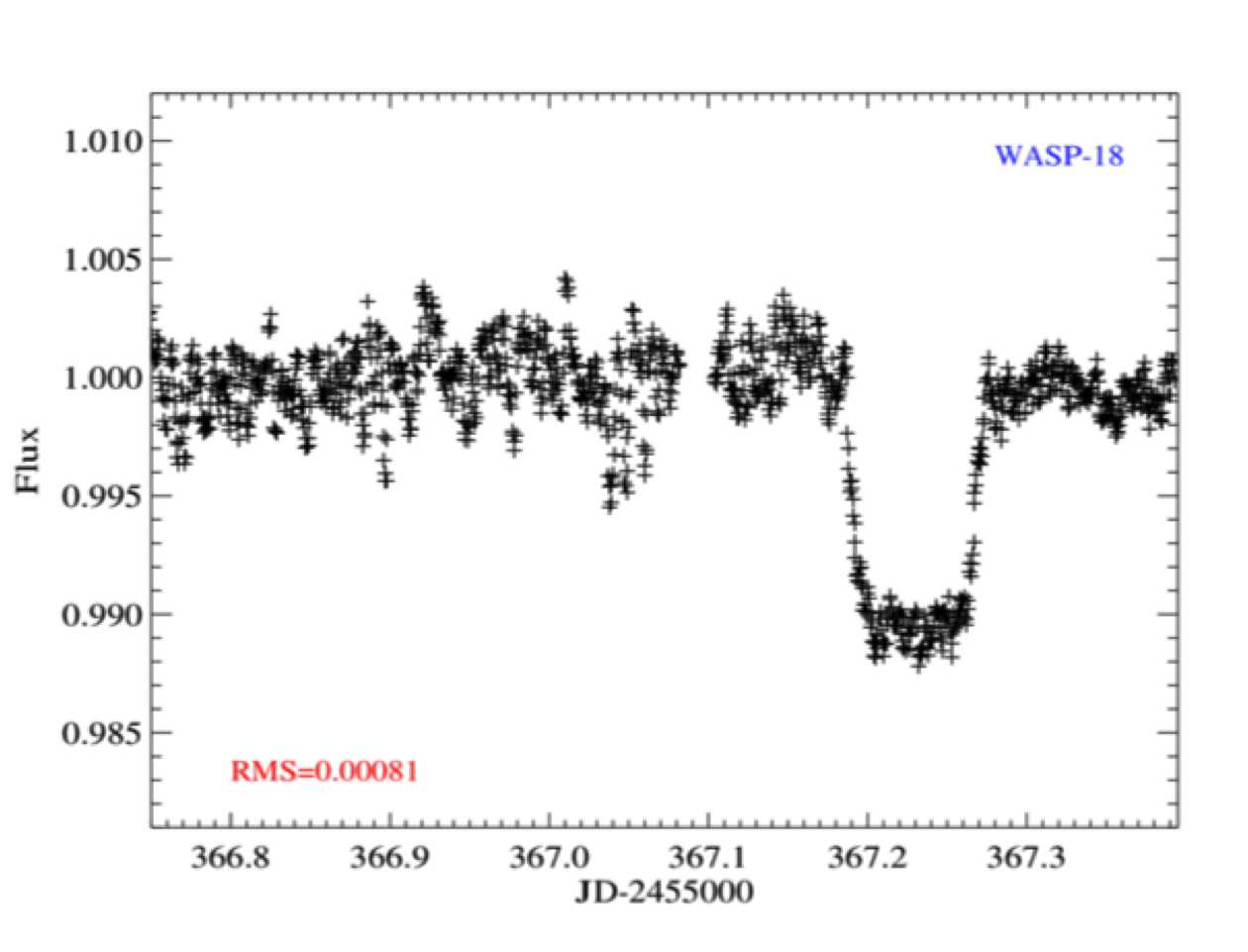}
 \caption[]{From Crouzet, 2010. Two photometric instruments have already been operated at Concordia. One, PAIX, is devoted to stellar variability and is focused on RR Lyrae stars displaying the still not unterstood Blazkho effect. Many good quality time series have been obtained during the winters 2009 and 2010. The second one is named A-STEP and is devoted to the search for exo-planet transits. After a prototype instrument started earlier, the 40-cm telescope has been successfully operated during the winter 2010. This is an example of a transit measured with a photometric accuracy better than 1 millimagnitude on a magnitude 9 star}
 \label{monobloc}
   \end{figure}         
    
     \begin{figure}[htbp]
\centering
 \includegraphics[width=12cm]{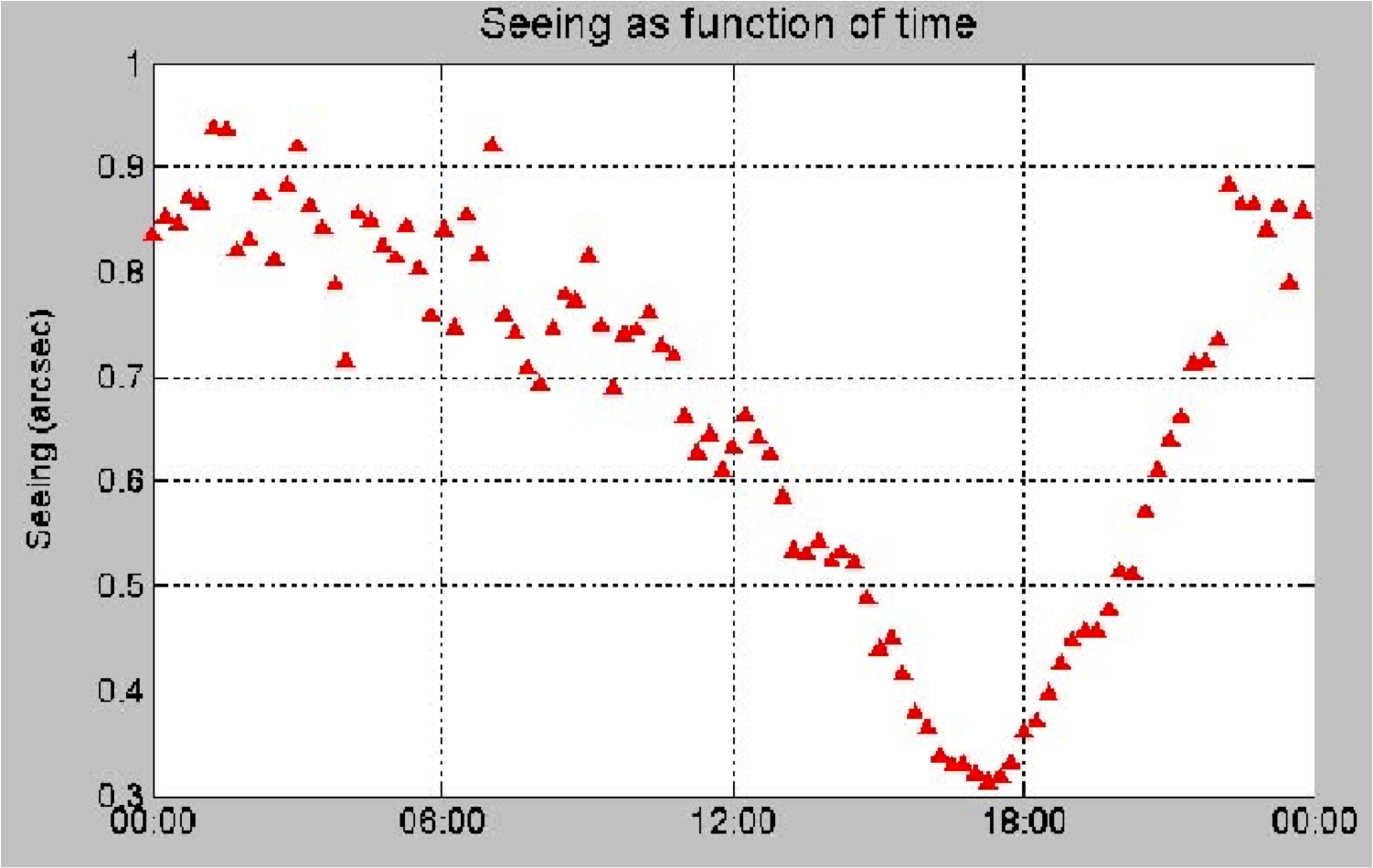}
 \caption[]{From Aristidi et al, 2007. The seeing was first measured during the summer seasons before the first winter-over year, starting in 2002/2003 on the star Canopus. It proved to be strongly time dependent, as this figure shows, with an excellent mean value of the order of 0.3 arc-sec during the local afternoons. This is the mean daily behavior averaged on the complete 2003/2004 summer season. This peculiarity is of course extremely promising for the future solar programs.}
 \label{monobloc}
   \end{figure}      
    
     \begin{figure}[htbp]
\centering
 \includegraphics[width=12cm]{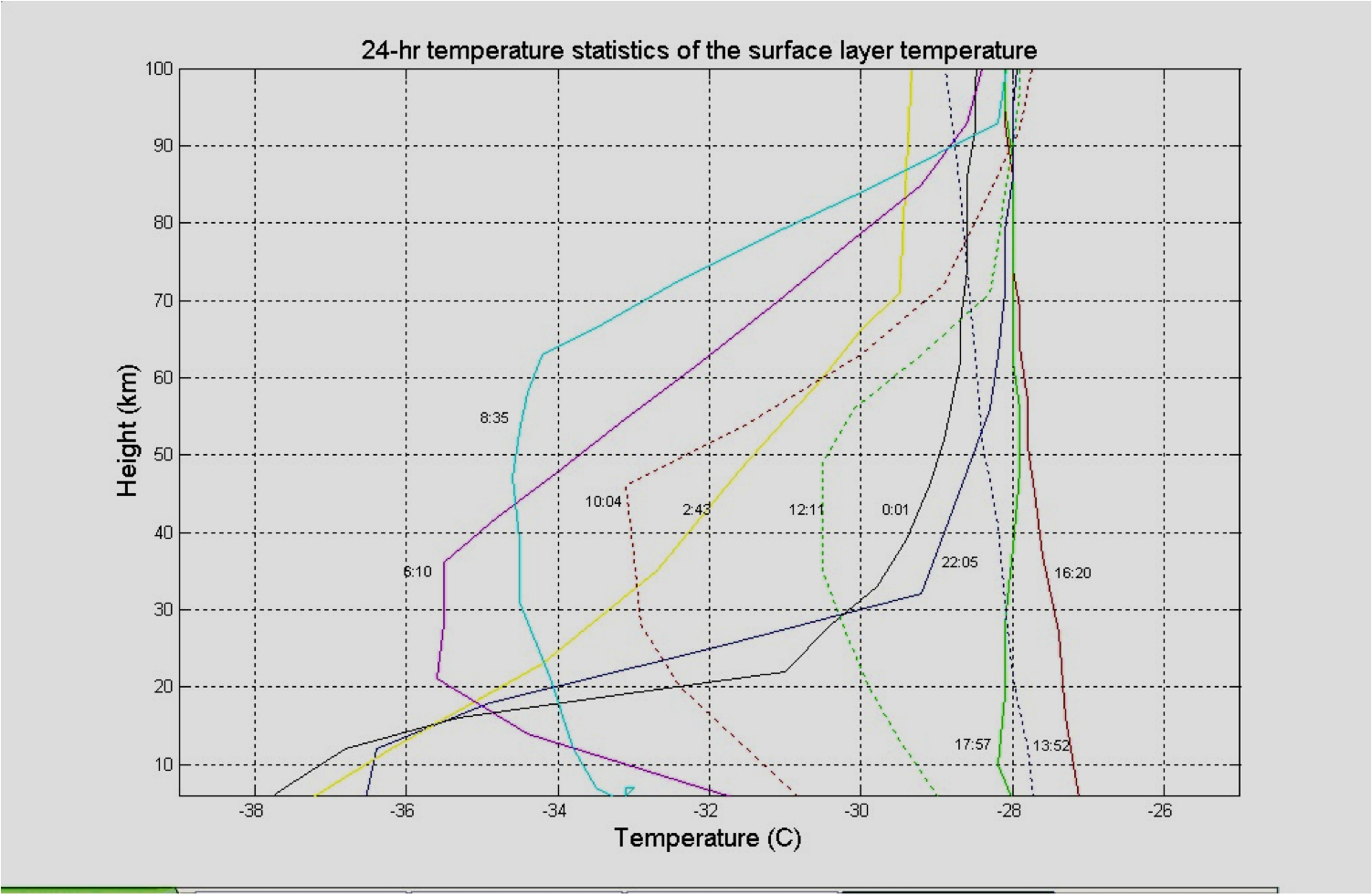}
 \caption[]{From Aristidi, 2006, Private comm.Ten meteorological radiosondes were launched during 24 hours  in January, 2004 to probe the temperature gradients inside the first 100 meters above the ice surface. In this summer situation, a strong inversion gradient of about 10C is visible at local midnight on the first 30 to 40 meters, that vanishes during the local afternoon, explaining the seeing behavior shown by the previous figure.}
 \label{monobloc}
   \end{figure}  
    
   \begin{figure}[htbp]
\centering
 \includegraphics[width=12cm]{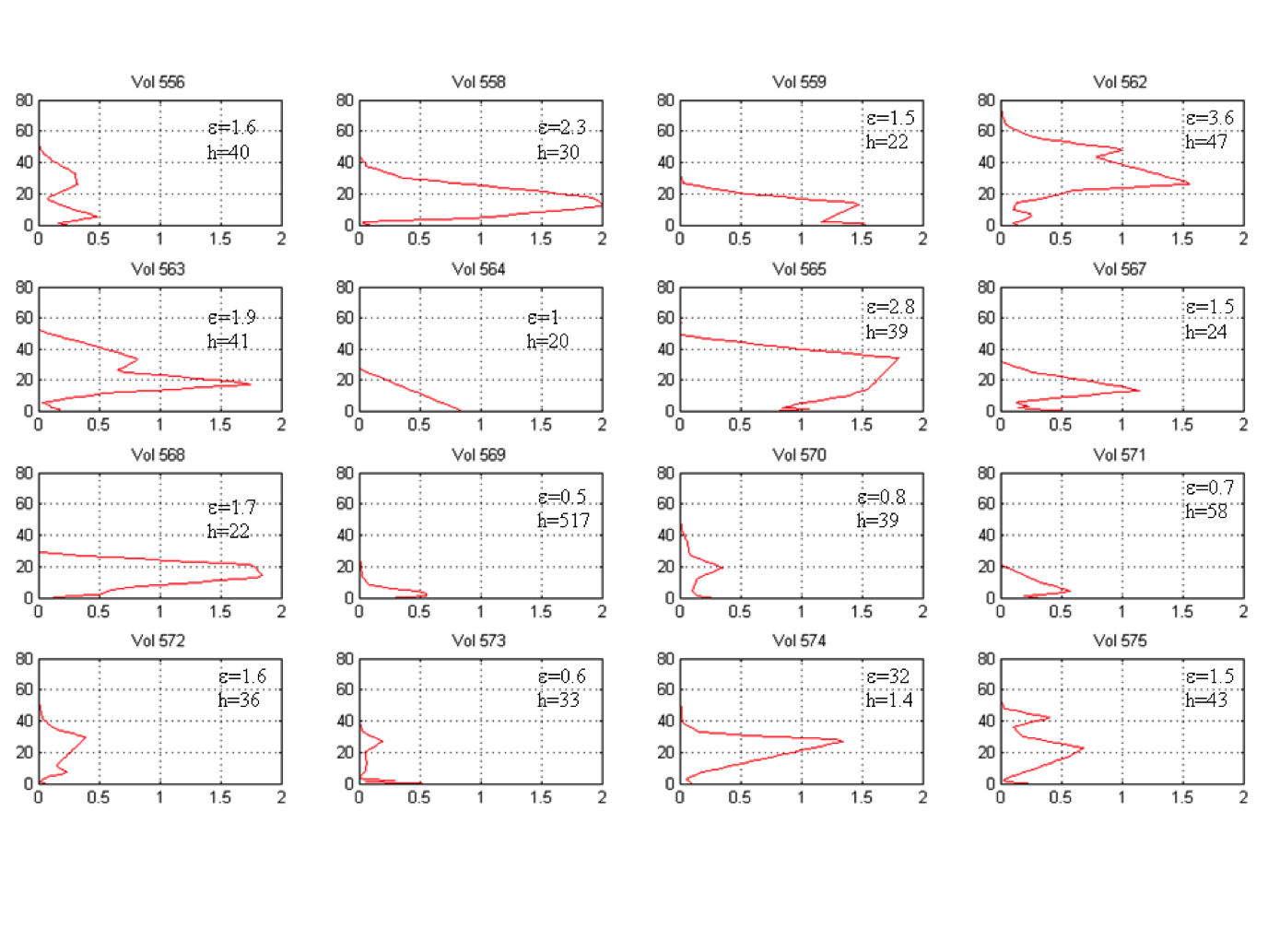}
 \caption[]{From Aristidi, 2006, Private comm. In winter when the Sun is totally absent, the midnight situation of the previous figure becomes permanent, with more amplitude. That creates a turbulent surface layer and being combined with a similar wind speed gradient, it becomes an optically turbulent layer of a few dozen meters of thickness. This shows 16 samples of the turbulent profile in the first 80 meters, in linear scale, measured by radiosounding with microthermal sensor equipment during the winter 2005.  }
 \label{monobloc}
   \end{figure}      

 \begin{figure}[htbp]
\centering
 \includegraphics[width=12cm]{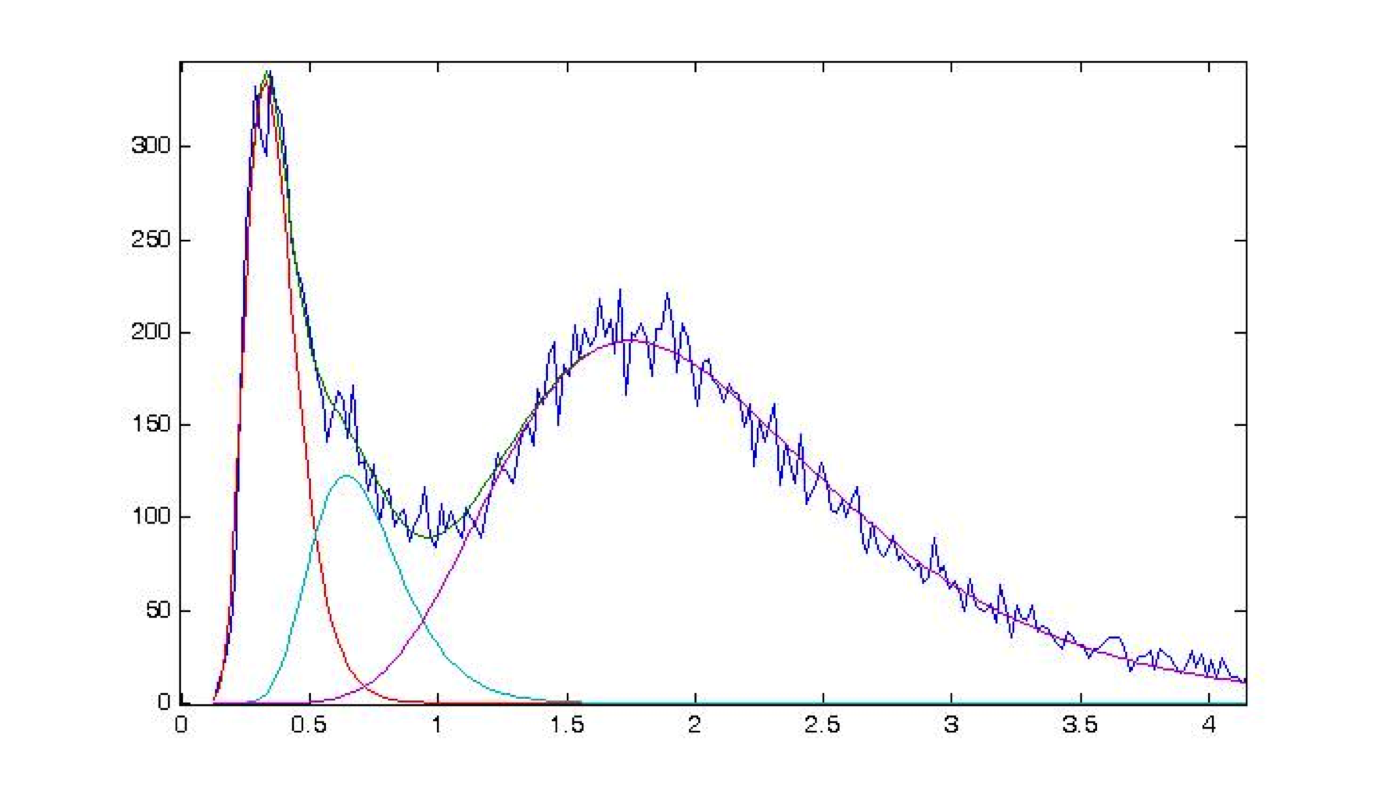}
 \caption[]{From Aristidi et al, 2009. Histogram of the winter seeing between 0 and 4 arc-sec, measured in 2006 by a DIMM  located between 7 and 8 meters above the surface on the Concordiastro platform. It looks nearly bimodal. The best fit is obtained by the sum of 3 log-normal functions, all three being independently shown here. The right part corresponds to situations when the telescope is embedded inside the surface turbulent layer. The left one when the telescope is outside this layer, virtually in the free atmosphere, and the middle one is the small fraction (10 per-cent) of intermediate situations, either due to the presence of a second fainter turbulent layer above the main one, or due to the random motion up and down of the sharp upper limit of the surface layer around the telescope entrance aperture during the two minutes integration time.}
 \label{monobloc}
   \end{figure}  

 \begin{figure}[htbp]
\centering
 \includegraphics[width=12cm]{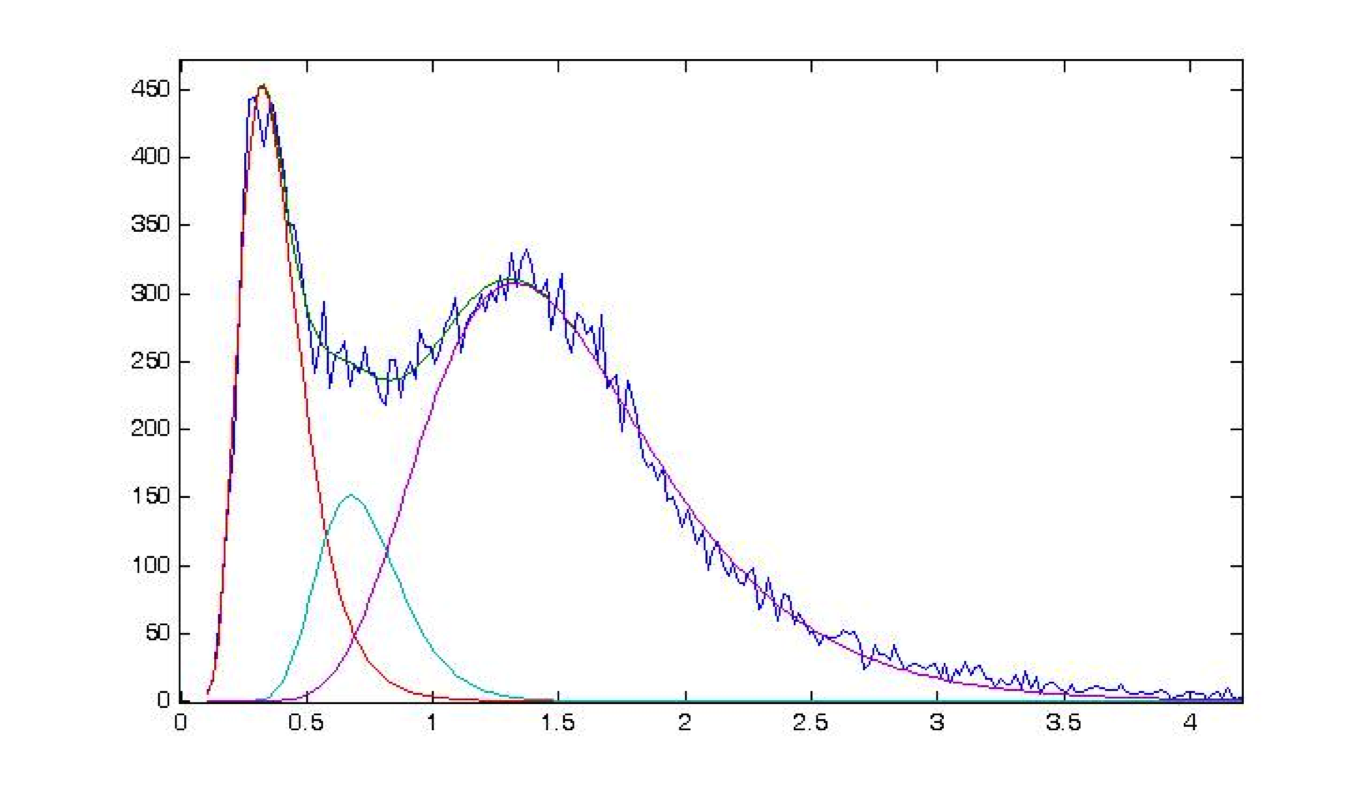}
 \caption[]{From Aristidi et al, 2009. During intermediate seasons, when the Sun is as anywhere else alternately present or absent (this is autumn, 2007),  the histogram looks very similar, with the only exception of showing less turbulence on the right curve, when the telescope is inside the surface layer. The other two fits are nearly identical to the winter figure, indicating that the free atmosphere seeing does not show any seasonal dependence}
 \label{monobloc}
   \end{figure}  

 \begin{figure}[htbp]
\centering
 \includegraphics[width=12cm]{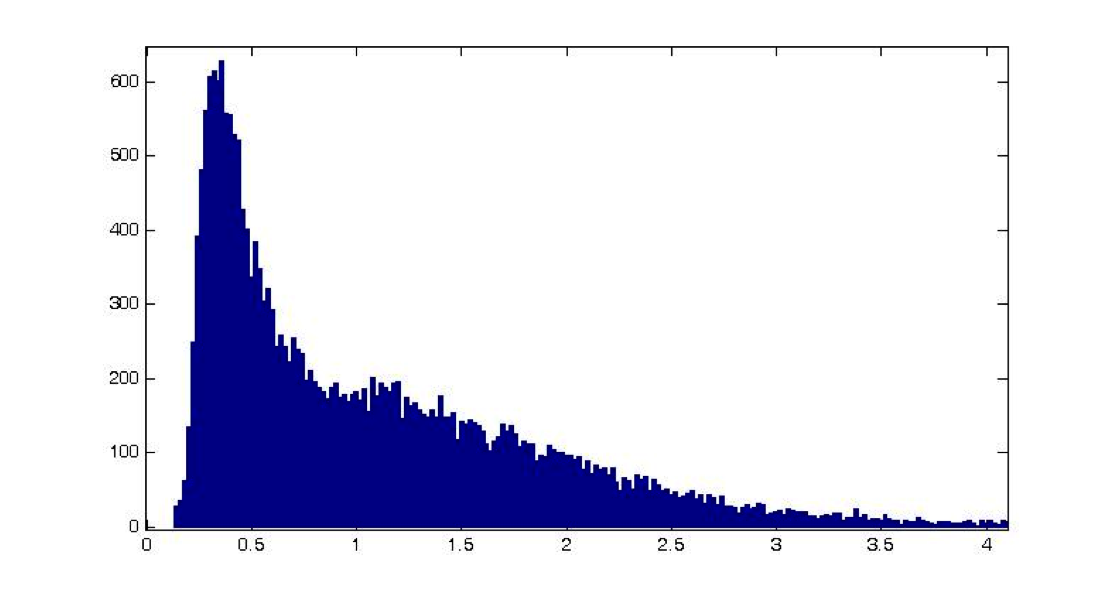}
 \caption[]{From Aristidi et al, 2009. In August, 2005, during the first winter-over, the GSM instrument, standing on the ice surface, failed. It is composed of two identical DIMMs and one of them remained operational. It was the personal initiative of K. Agabi to move it on the roof of the Concordia station, at 20m high, where 3 and half months of data could then be recorded, half in Winter and half in Spring.  The histogram of this 20m high data set shows that the distribution of the "free atmosphere seeing" remains identical, around 0.3 arc-sec, but becomes just more frequent, around 45 percent of the time against less than 20 percent at 7-8 m}
 \label{monobloc}
   \end{figure}  

 \begin{figure}[htbp]
\centering
 \includegraphics[width=12cm]{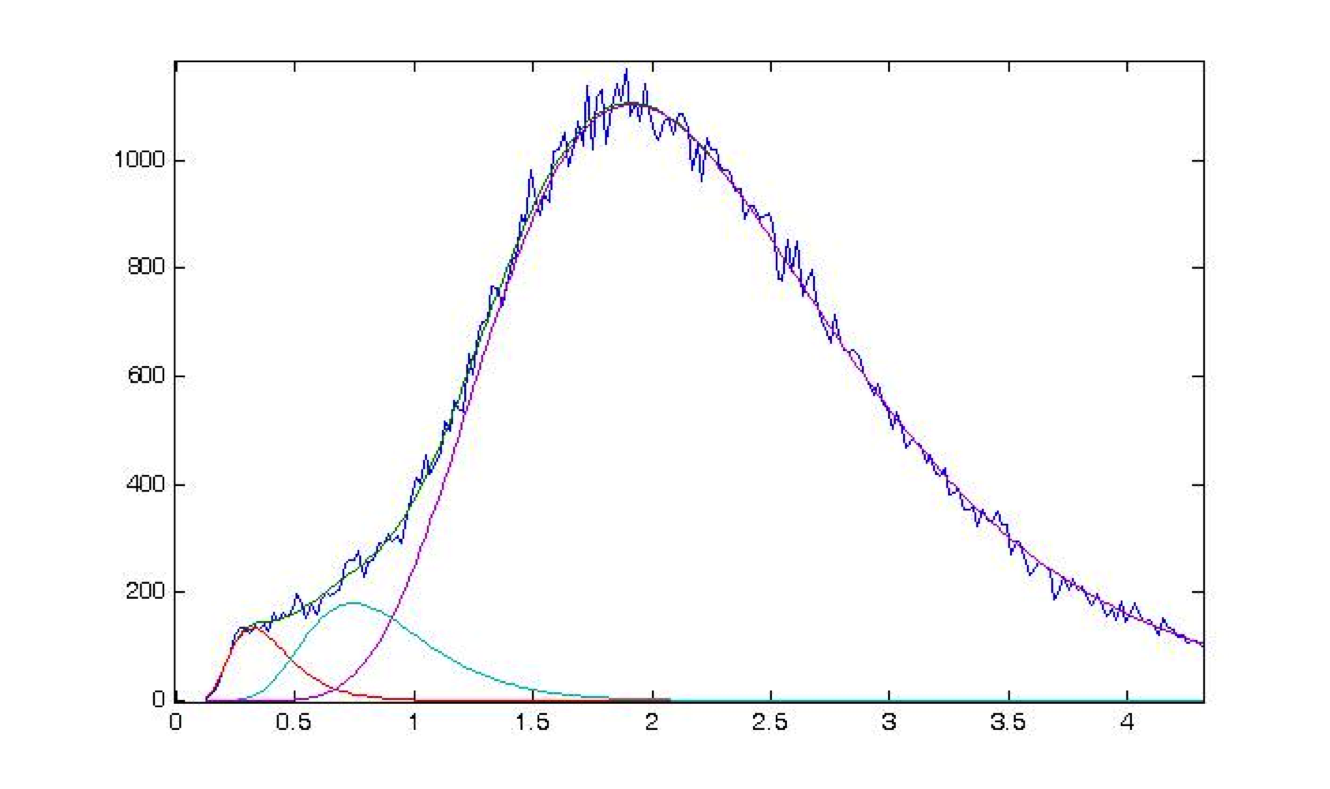}
 \caption[]{From Aristidi et al, 2009. The same histrogram obtained from the GSM telescopes standing on the snow surface, their apertures being at about 2 m high. It shows a small but still clearly visible fraction of free atmosphere seeing, of the order of 2 percent of the time on this  2005+2006 winter histogram. Apart from this very small probability of occurrence, this free atmosphere seeing distribution is still exactly the same, around 0.3 arc-sec, providing another evidence for the fact that above the sharp upper limit of the turbulent surface layer, there is nothing else than the free atmosphere, even when this upper limit is located lower than 2 meters of altitude.}
 \label{monobloc}
   \end{figure}  

\begin{figure}[htbp]
\centering
 \includegraphics[width=12cm]{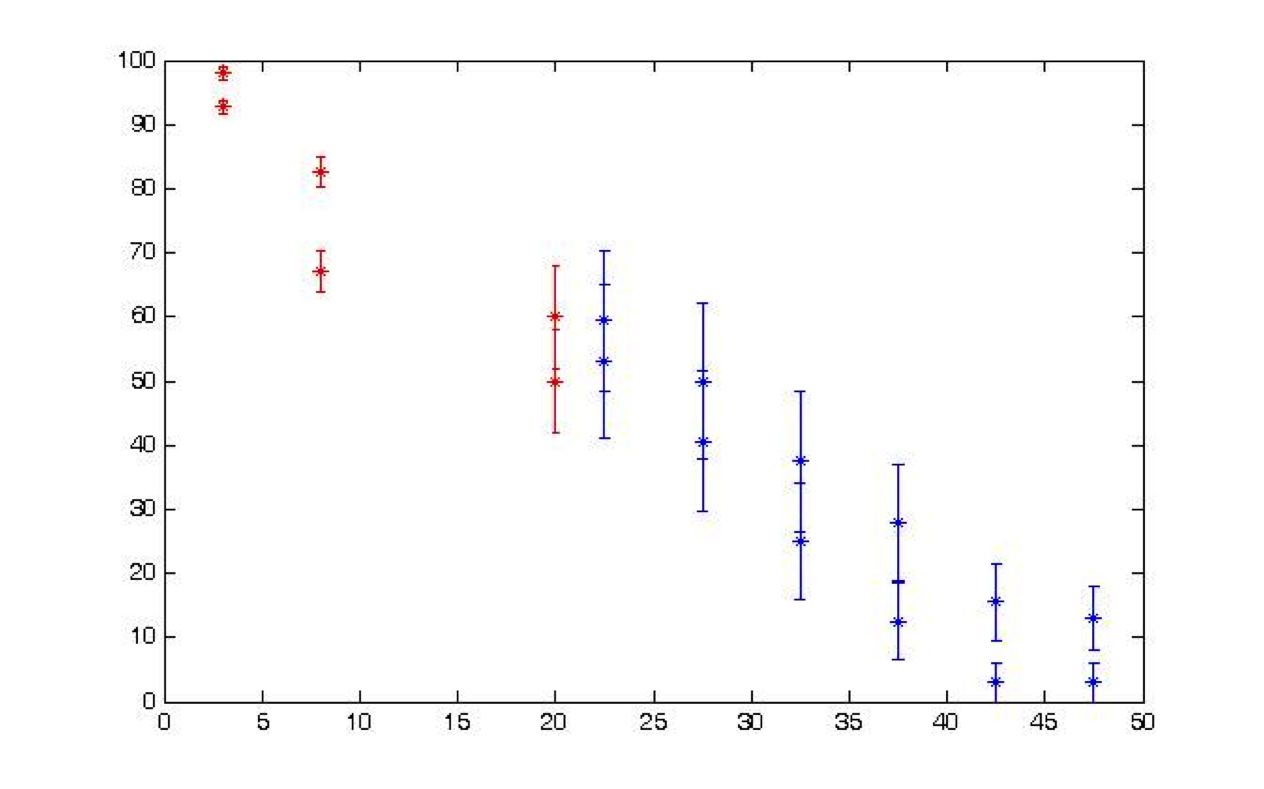}
 \caption[]{From Aristidi et al, 2009. Adding up the statistical informations provided by the three DIMMs located at 3 different heights and the 32 snapshop informations provided by the 32 available radiosoundings during the winter 2005, it becomes possible to plot the probability of being inside (or outside) the surface layer as a function of height.  Two values are plotted at each height. The upper one corresponds to the left part of the histogram, meaning being totally in the free atmosphere (2 percent probability at 2m high for instance, or 98 percent not being outside the surface layer), the other one meaning being outside the main turbulent layer, with the possibility of still having some secondary faint turbulent layer somewhere above.  Altogether, the median value of the surface layer thickness can be estimated between 25 and 30 meters. Other key numbers are the fraction of turbulent energy inside the surface layer, 95 percent, and the fraction of this turbulent energy statistically remaining above the height of 25 m, only 25 percent.}
 \label{monobloc}
   \end{figure}  

\begin{figure}[htbp]
\centering
 \includegraphics[width=10.5cm]{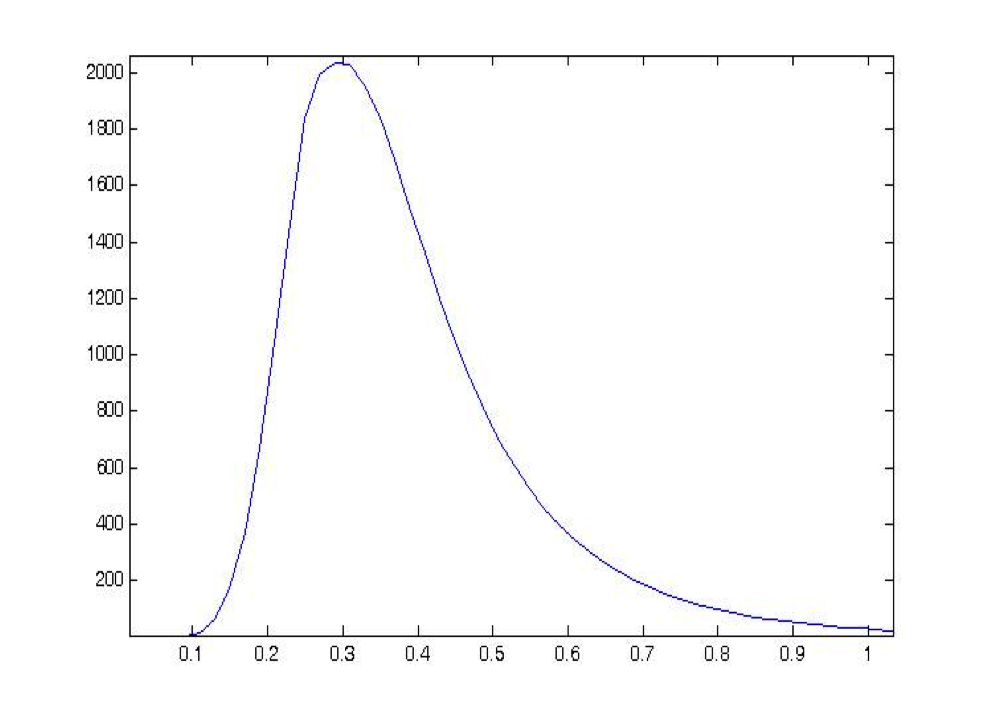}
 \caption[]{From Aristidi et al, 2009. Free atmosphere seeing distribution, from 0 to 1 arc-sec. This curve is the addition of all left parts of the log-normal fits on the histograms. 30 months of data have been summed, from winters, springs and autumns just excluding the summers. The peak value is at 0.29 arc-sec with a median value at 0.34 This distribution appears to be independent of height provided that the measurement is made outside the surface layer. The agreement with a recent model of the Forot group (Lascaux et al, 2010) is quite remarkable. The median altitude of this surface layer stands then between 25 and 30 m while the first quartile can be estimated to be between 12 and 15 m from the previous figure. }
 \label{monobloc}
\end{figure}  

\begin{figure}[htbp]
\centering
 \includegraphics[width=10.cm]{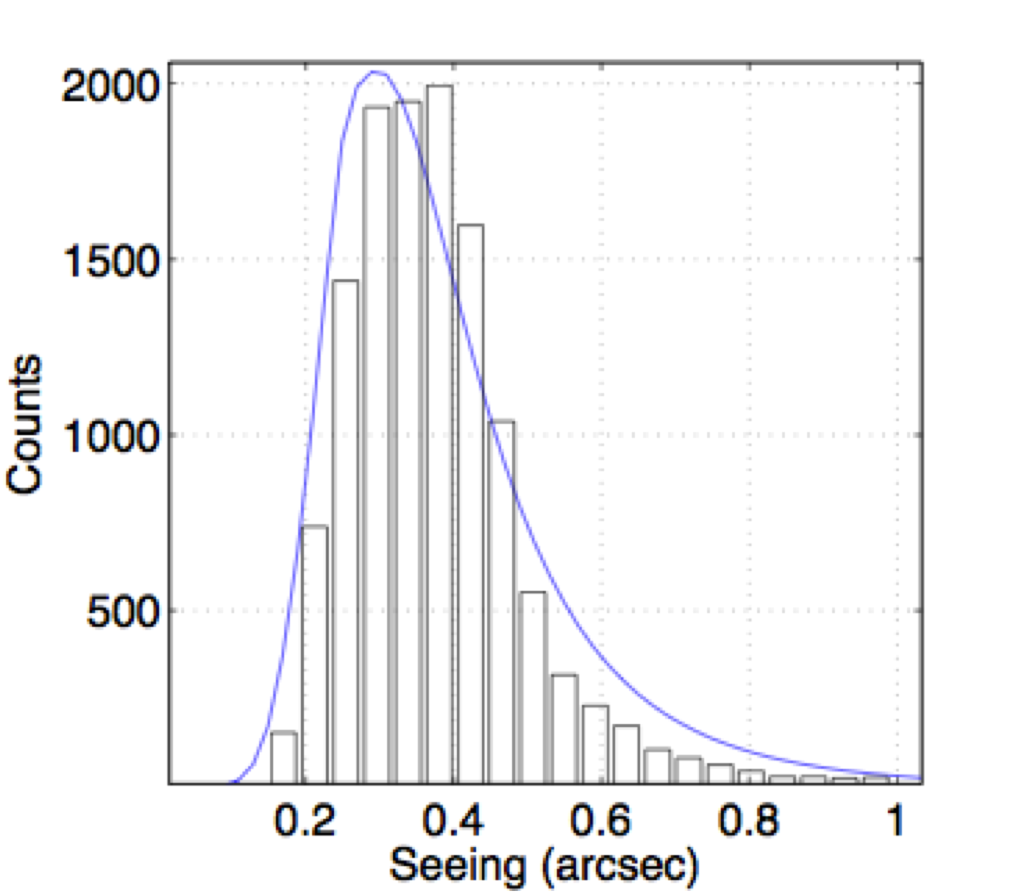}
 \caption[]{From Aristidi et al, 2009. Superimposed on the free atmosphere seeing distribution presented in the previous figure is shown the histogram of the summer seeing strictly limited to one hour of data centered each day on the best afternoon value. The agreement between the two is remarkable and is a convincing evidence for the absence of anything else than the free atmosphere above the surface layer, since when this turbulent surface layer vanishes in summer, the seeing is back to its winter values obtained above it. This free atmosphere seeing is then found independent of the season, including summer.}
 \label{monobloc}
\end{figure}  

\begin{figure}[htbp]
\centering
 \includegraphics[width=10.5cm]{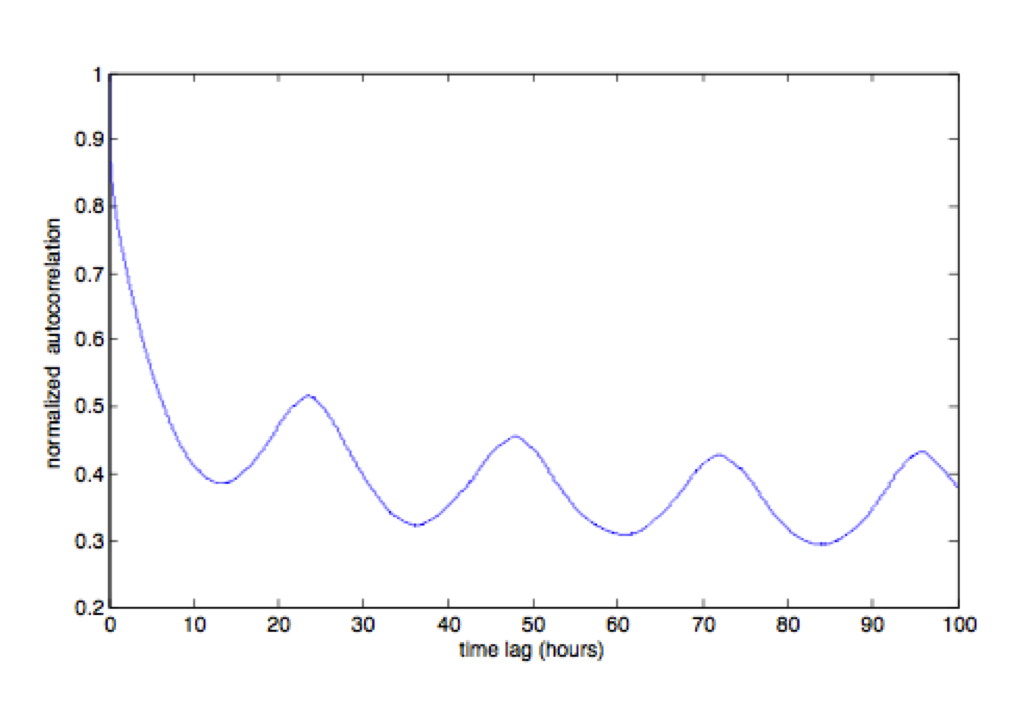}
 \caption[]{From Fossat et al, 2010. The next interesting question deals with the temporal distribution of the good seeing episodes. It is not a trivial question to answer as the DIMM data sets are not continuous, being often interrupted by adverse weather or instrumental problems. Only a statistical study can then be considered, by means of the autocorrelation method. This has been done on the data from the DIMM located on the Concordiastro platform (between 7 and 8 m), since it is by far the longest data set available, nearly 5 years. From any data set, two binary files are created, the first one for the existing data, 1 if data exists, 0 if not, the second one for the good seeing data, 1 if the measured seeing is better that a given threshold, 0 if it is worse. The autocorrelations of these two binary files are then exploited to study the statistical behavior of the good seeing episodes. The autocorrelation shown here has been processed on the window function of the two winters 2005 and 2006, when there was only one winter-over astronomer in charge. The resulting 24-hour periodicity is clear and demonstrates that the DIMM was not totally automated yet. The "winter" here and in the rest of the presentation covers 6 months from the end of March to the end of September. They are the 6 cold months without significant temperature drifts, see Fig. 5. } 
 \label{monobloc}
 \end{figure}  

\begin{figure}[htbp]
\centering
 \includegraphics[width=11cm]{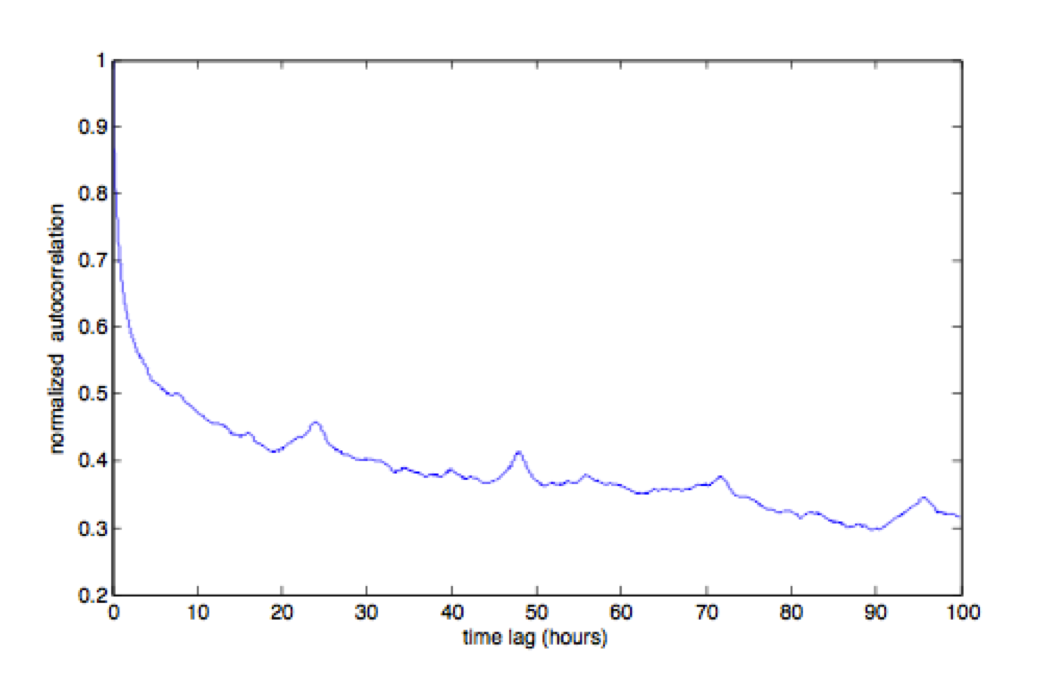}
 \caption[]{From Fossat et al, 2010. Same as previous figure for the window function of the winter 2007 when for the first time, two astronomers were present and could share the duty. The 24-hour periodicity of the window function is not totally suppressed but it is significantly reduced. } 
 \label{monobloc}
\end{figure}

\FloatBarrier

\begin{figure}[htbp]
\centering
 \includegraphics[width=9cm]{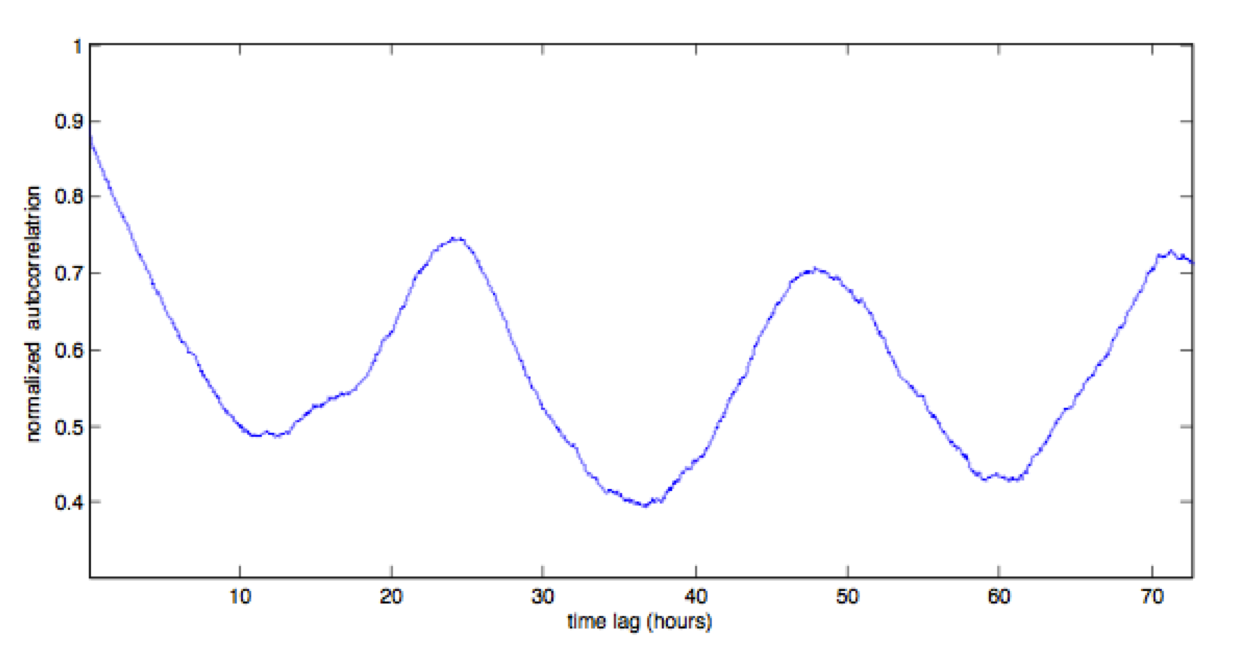}
 \caption[]{From Fossat et al, 2010.   By dividing the autocorrelation of the existing good seeing window by the autocorrelation of the existing window function, the result is the true autocorrelation of the good seeing windows. This example is the summer good seeing windows autocorrelation, using a threshold of 0.5 arc-sec for the definition of "good seeing". It shows the strong 24-hour periodicity of the seeing (see Fig. 7; now it is no more a window function periodicity). It deserves several comments: First the 0.25 linear decrease in the first 6 hours or so corresponds to the daily window of excellent seeing from 14: to 20: local time, and thus confirms the efficiency of this analysis. On the other hand, the asymptotic 24-hour oscillation  around a value of the order of 0.55 indicates that in summer, 55 percent of the measurements are better than 0.5 arc-sec. That is by far much more than 6 hours per day and says that during the remaining 18 hours every day, there is still a large fraction of measurements better than this threshold of 0.5 arc-sec, they are just more randomly distributed. A last comment is the quick drop from 1 to 0.88 at the beginning, that indicates a number of about 12 percent of individual good values in a moment of bad seeing, or bad values in a moment of good seeing (just uncorrelated with the close neighboring).}
 \label{monobloc}
\end{figure}     

   \FloatBarrier
  
   \section*{}

\begin{figure}[htbp]
\centering
 \includegraphics[width=9cm]{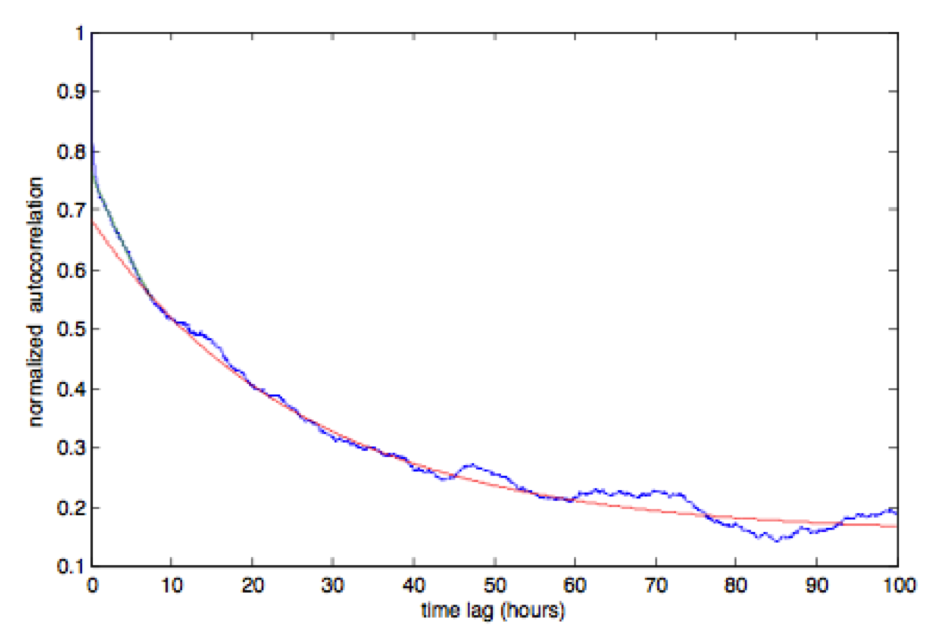}
 \caption[]{From Fossat et al, 2010. Autocorrelation of the good seeing (better than 0.5 arc-sec) window function averaged on 4 winters (24 months), and thus statistically quite robust. Comments: 1. The 24-hour periodicity has essentially disappeared as the Sun is absent most of the defined winter. 2. The initial drop shows the existence of 20 percent of good seeing during a bad seeing moment or of bad seeing during a good seeing moment.  3. The linear drop during the 7-8 first hours reminds the summer situation and indicates that the good seeing episodes are mostly never shorter than 7 or 8 hours, with the exception of a fraction of individual values.  4.  The asymptotic value slightly less than 20 percent is of the order of the probability of being above the surface layer at 8m (see Fig. 14).  5.  In between, the exponential decrease with an e-damping time of about 30 hours can find a meteorological interpretation. See Fossat et al, 2010 for more details.} 
 \label{monobloc}
\end{figure}  
   \section*{} 
   \FloatBarrier

   \begin{figure}[htbp]
\centering
 \includegraphics[width=11cm]{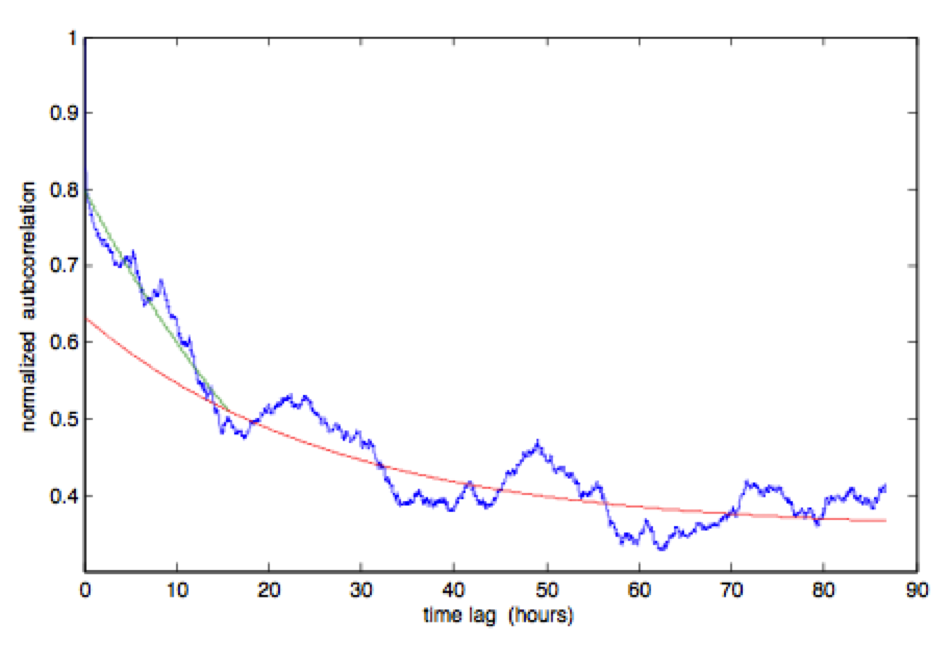}
 \caption[]{From Fossat et al, 2010. In 2005, a DIMM was operated during 3 and half months at 20m high on the roof of the Concordia station (see Fig. 12). This is the autocorrelation of the good seeing window function from this 20m high data set.  With less statistical robustness, the main features of the previous  figure are still present. The fraction of isolated good or bad seeing values remains at 20 percent, the 30-hour exponential damping remains unchanged, but the initial linear decrease lasts now 15 hours and the asymptotic value is close to 40 percent. A mathematical modeling of the good seeing occurrences has been adjusted to reproduce this general autocorrelation. In includes a Poisson distributed recurrence of good episodes with a typical delay of 5 days, each one including about two days filled at more than 80 percent of good seeing values, with individual sequences not shorter than 15 hours. Note that the 24-hour periodicity is visible again, as this 20m height data set was obtained only partly in winter and partly in springtime. } 
 \label{monobloc}
\end{figure}   

  \begin{figure}[htbp]
\centering
 \includegraphics[width=11cm]{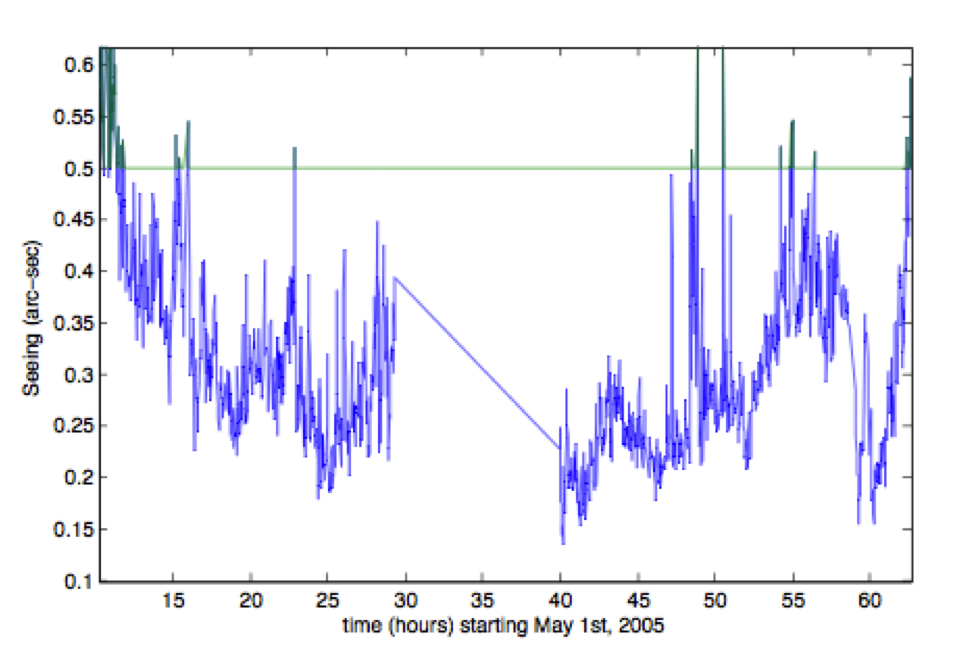}
 \caption[]{From Fossat et al, 2010. An example of two and half days of seeing measured at 8m high, including an 11-hour gap. The mean seeing value is 0.29 arc-sec during 60 hours.} 
 \label{monobloc}
\end{figure}   

 \begin{figure}[htbp]
\centering
 \includegraphics[width=11cm]{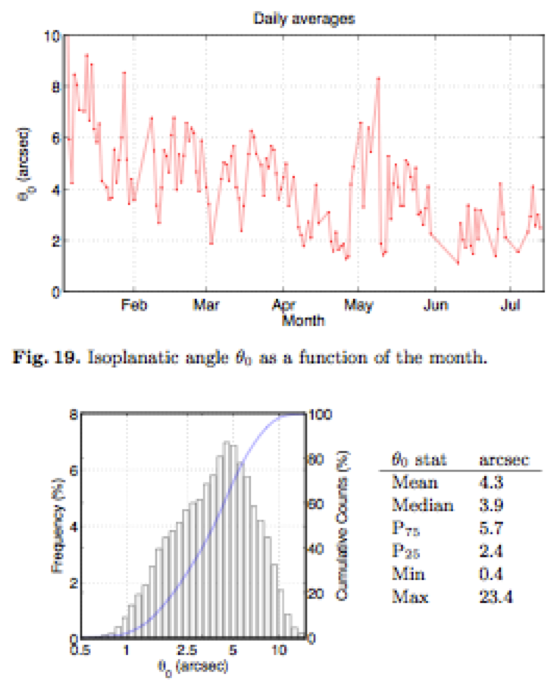}
 \caption[]{From Aristidi et al, 2009. The isoplanatic angle has been measured during a little longer than 6 months in 2006, from middle summer to middle winter. Its histogram and mean values are definitely better than anywhere else, but it shows a clear seasonal dependence, being at its best in summer and worsening during the coldest season.} 
 \label{monobloc}
\end{figure}   
  
  \begin{figure}[htbp]
\centering
 \includegraphics[width=14cm]{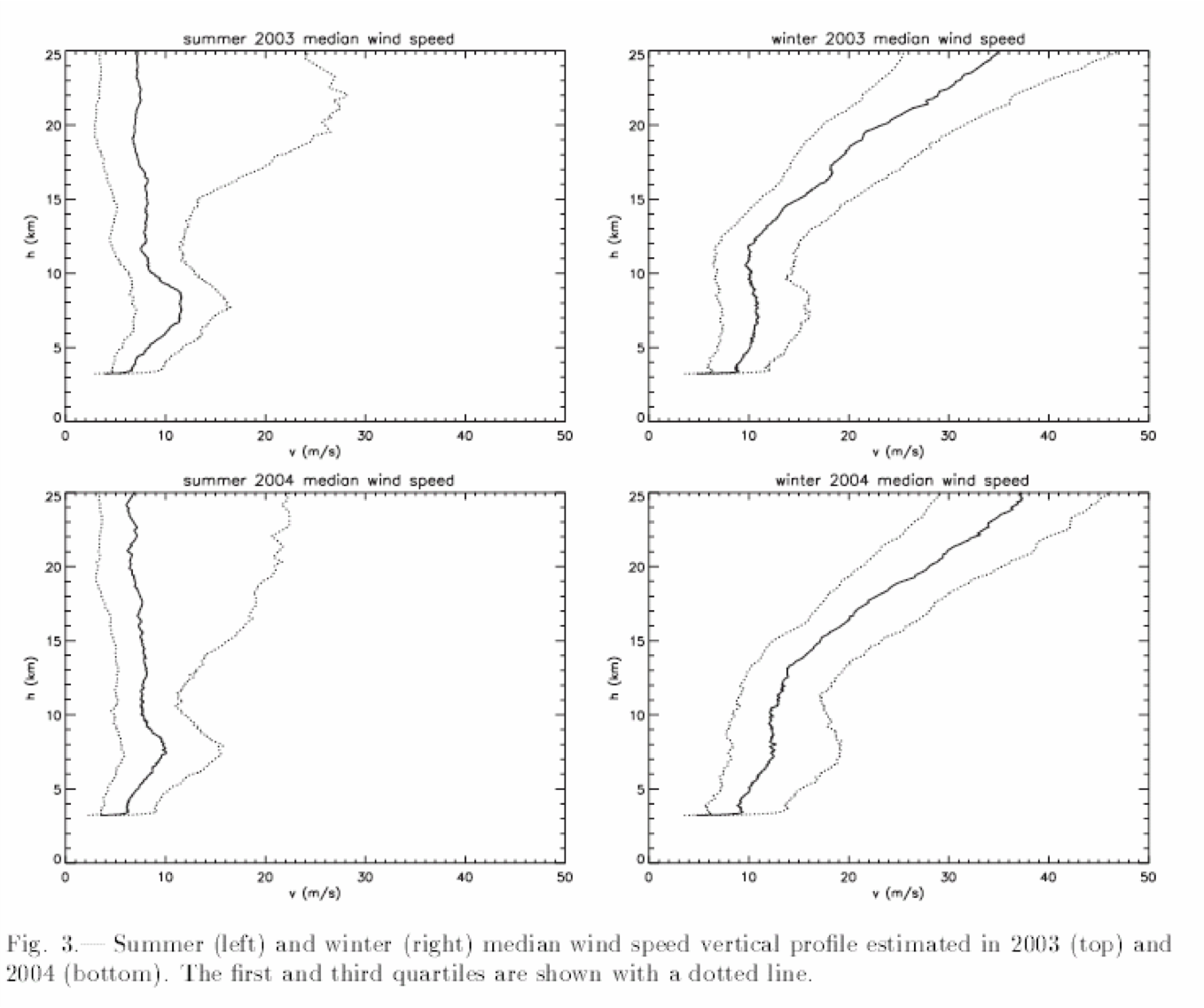}
 \caption[]{From Geissler and Masciadri, 2006. The vertical wind profiles above Dome C are quite different in summer and winter, certainly explaining the seasonal dependence of the isoplanatic angle which strongly depends on the high altitude winds.} 
 \label{monobloc}
\end{figure}

  \section{Conclusion}

This presentation was just a summary of the main assets of the Dome C as an astronomical site, with some emphasis given on the peculiarities of the seeing and its temporal behavior. And a little more. It is certainly to be regarded as one of the best astronomical sites in the world in general, and the best one on a few scientific niches exploiting for instance its infra-red and its high angular resolution properties, or its unique daytime seeing and coronal sky. Only the other sites on the Antarctica high plateau, around the Dome A, can presumably rival, being potentially even better regarding the turbulence surface layer properties. On the other hand, they will suffer from more pollution by the austral auroras, from which the Dome C is relatively well protected thanks to its geographical location not far from the magnetic Pole. And of course, the Dome C has the important advantage of the existence of the Concordia winter-over station, and more than 10 years of experience in its logistics by the Italian and French Polar operators, PNRA and IPEV. Despite this experience, it is clear that it will always be more difficult to operate any important astronomical instrumentation at Dome D than it could be in a more temperate site. As more difficult automatically implies more expensive, any ambitious project needs a serious and complete study of its feasibility and its additional costs. By now, the relatively modest photometric instruments in operation are expected to produce good science and to promote the site. They could be followed by a spectroscopic night time instrument, and/or a specific solar project before the implementation of a 2 to 3-m infra-red telescope (PLT) that seems to be an essential step before any more ambitious project can be attempted.


\begin{thebibliography}{}

 
  \bibitem[2005]{aris05} Aristidi E. et al., 2005, A\&\/A 444, 651
  \bibitem[2009]{aris09} Aristidi E. et al., 2009, A\&\/A 499, 955

   \bibitem[2006]{tomasi06} Tomasi, C., Petkov, B., Benedetti, E., Vitale, V., Pellegrini, A., Dargaud, G., De Silvestri, L., Grigioni, P., Fossat, E., Roth, W. and Valenziano, L.  2006, J.G.R., 111. 
    
   \bibitem[2010]{crouzet10} Crouzet, N., 2010, PhD Thesis, Universite de Nice
   
    \bibitem[2007]{geissler06} Geißler K., and  Masciadri E., 2006, PASP, 118, 845, 1048   
    
    \bibitem[2007]{mosser07} Mosser, B. and Aristidi, E.  2007, PASP, 119, 127
    
    \bibitem[2010]{fossat10} Fossat, E., Aristidi, E., Agabi, K., Bondoux, E., Challita, Z., Jeanneaux, F. and Mekarnia, D.  2010, A\&\/A 517, 69F
    
     \bibitem[2010]{Lascaux10} Lascaux, F., Masciadri, E., Hagelin, S., 2010, MNRAS, in press
         
    \end{thebibliography}
\end{document}